\documentclass[prl,twocolumn]{revtex4}%
\usepackage{amsfonts}
\usepackage{amsmath}
\usepackage{amssymb}
\usepackage{graphicx}%
\providecommand{\U}[1]{\protect\rule{.1in}{.1in}}
\begin{document}
\preprint{cond-mat/0612135 }
\title{Equilibrium topology of the intermediate state in type-I superconductors of
different shapes}
\author{Ruslan Prozorov}
\email{prozorov@ameslab.gov}
\affiliation{Ames Laboratory and Department of Physics \& Astronomy, Iowa State University,
Ames, Iowa 50011}

\pacs{PACs:}

\begin{abstract}
High-resolution magneto-optical technique was used to analyze flux patterns in
the intermediate state of bulk Pb samples of various shapes - cones,
hemispheres and discs. Combined with the measurements of macroscopic
magnetization these results allowed studying the effect of bulk pinning and
geometric barrier on the equilibrium structure of the intermediate state.
Zero-bulk pinning discs and slabs show hysteretic behavior due to topological
hysteresis -- flux tubes on penetration and lamellae on flux exit.
(Hemi)spheres and cones do not have geometric barrier and show no hysteresis
with flux tubes dominating the intermediate field region in both regimes. It
is concluded that flux tubes represent the equilibrium topology of the
intermediate state and that the laminar structure is unstable towards Lorentz
or condensation energy forces. Real-time video is available in \cite{video}.

\end{abstract}
\date{November 2006}
\maketitle

Pattern formation in strongly correlated systems is a topic of incessant
interest for a broad scientific community \cite{walgraef1997}. At a first
sight, type-I superconductors represent a perfect physical system where it is
relatively easy to tune the parameters and try to understand the physics
behind observed topology of the intermediate state. In fact, analogies to
type-I superconductors extend from astrophysics \cite{buckley2004} to the
physics of ice \cite{wang2006}. The fundamental problem, however, is that in a
finite system it is impossible to predict 2D and moreover 3D pattern based
solely on the energy minimization arguments \cite{choksi2004}. The pattern has
to be assumed and then its geometrical parameters are determined from the
minimization. Back in the 1930s Lev Landau suggested a simple stripe model,
which was possible to analyze analytically \cite{landau1937,landau1943}. Later
refinements (such as domain widening and/or branching) tried to address
apparent inconsistencies between the model and the experiment
\cite{livingston1969,huebener2001,tinkham2004}. Still a comprehensive
description has never been achieved with the main problem being multiple
observations of the closed-topology structures (flux tubes) in best samples.
Experimental and theoretical effort is growing to obtain general understanding
of the problem \cite{liu2001,goldstein1996,dorsey1998,choksi2004}. Here we
outline several factors that influence and often determine the topology of the
intermediate state and must be taken into account by a successful theory.%

\begin{figure}
[ptb]
\begin{center}
\includegraphics[
height=2.6333in,
width=3.1548in
]%
{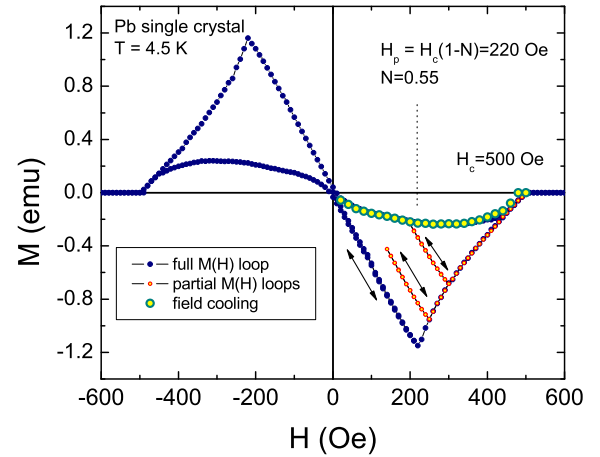}%
\caption{$M(H)$ loop measured in a Pb single crystal at $T=4.5$ K. Larger open
symbols show measurements each after field cooling at indicated field to $4.5$
K. Small open symbold show minor hysteresis loops.}%
\label{xtalMH}%
\end{center}
\end{figure}

First issue is (bulk) flux pinning. In the overwhelming majority of published
papers this question is simply omitted. We have shown that tubular topology is
fairly robust, but is destroyed and transformed into a laminar pattern by the
structural defects. In more disordered samples with significant bulk pinning,
a non-equilibrium dendrite - like topology of the intermediate state is
observed \cite{prozorov2005}. Recently, it was shown that close to $H_{c}$
laminar structure transforms into tubular upon applying a small-amplitude AC
field to shake it out of the metastable state \cite{menghini2007}.
Fortunately, it is easy to distinguish between pinning-induced and topological
hysteresis. The former becomes larger at the lower fields and reaches maximum
at $H=0$. The latter is maximal around $H=H_{c}\left(  1-N\right)  $ (where
$N$ is the demagnetization factor) and should vanish in the limit of $H=0$. We
note that a concept of topological hysteresis as applied to type-I
superconductors was introduced to describe the irreversibility in a
\emph{macroscopic} response due to different topologies of the intermediate
state, e.g., sample magnetic moment \cite{prozorov2005}. It was later used to
describe a subtle specific issue of a crossover between tubes and laminae in
films near the $H_{c}$ \cite{gourdon2006}.%

\begin{figure}
[ptb]
\begin{center}
\includegraphics[
height=13.294cm,
width=9.0655cm
]%
{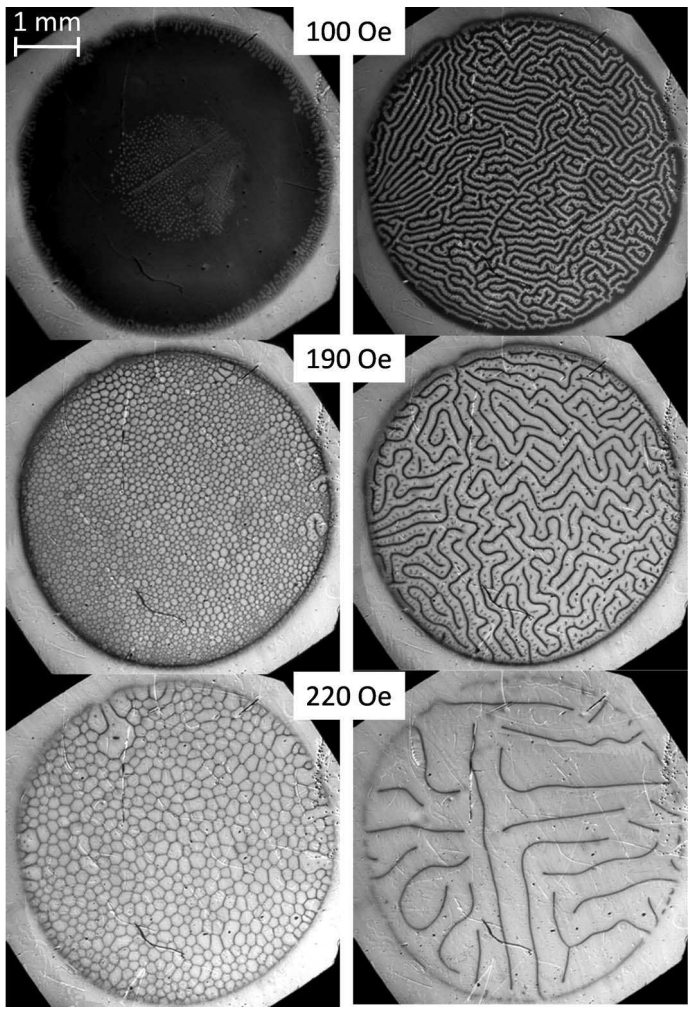}%
\caption{Structure of the intermediate state in a disc - shaped Pb single
crystal at \ $5$ K. Left column - increasing magnetic field after ZFC. Right
column - decreasing field.}%
\label{xtal}%
\end{center}
\end{figure}

Second crucial ingredient to understand the intermediate state is sample shape
and geometry. Films are relatively easy to make and work with, but they
provide only very limited experimental information and are at the borderline
of applicability of most theories. The effective penetration depth depends on
film thickness and, in case of Pb, the sample exhibits behavior of a type-II
superconductor below about $0.1$ $\mu m$ \cite{tinkham2004}. The phase
transition in the vicinity of $H_{c}$ is modified compared to the bulk case
\cite{abreu2004} and huge demagnetization factor causes any small defects on
the film edge to act as centers of premature flux penetration. In addition,
geometric barrier \cite{zeldov1994} plays dominant role and actually
determines the flux pattern \cite{castro1999}. On the other hand, experiments
with thick pinning-free type-I superconducting slabs and disks have
consistently showed the tubular pattern upon flux penetration and laminar
structure pattern upon flux exit. It was directly observed in single crystals
of Sn back in 1958 \cite{schawlow1958}, Hg \cite{huebener2001} and Pb
\cite{huebener2001,video}. The tubular pattern was reproduced numerically
\cite{bokil1997} and metastability of slabs was theoretically analyzed
\cite{fortini1980}. Ginzburg - Landau equations have stable multiquanta
solution for arbitrary tube size \cite{tinkham2004}. Still, the unsatisfactory
fact is that in any shape where there are two parallel surfaces perpendicular
to the magnetic field, there will be a geometric barrier that promotes an edge
instability and drives the tubes into the interior \cite{livingston1969}. The
barrier is different for flux exit, therefore tubes were considered to be a
result of the metastable state determined by the geometric barrier.

In this paper we study \textit{pinning-free} samples of various shapes -
discs, (hemi) spheres and cones. The latter two geometries do not have the
geometric barrier (as directly proven by the observations), yet show that flux
tubes exist both upon flux entry and exit proving that flux tubes represent
the equilibrium topology of type-I superconductors.

\textit{Quantum Design} MPMS magnetometer was used for magnetization
measurements. Magneto-optical (MO) imaging was performed in a pumped flow-type
optical $^{4}$He cryostat using Faraday rotation of a polarized light in Bi -
doped iron-garnet films with in-plane magnetization \cite{prozorov2005}. In
all images bright regions correspond to the normal state and dark regions to
the superconducting state. Due to very large volume of images and video, we
could only include a small subset of data in this paper. For a complete
coverage, including real-time video, see Ref.\cite{video}.

We begin with a single crystal of lead in form of a disk of diameter $d=5$ mm
and thickness $t=1$ mm. Four crystals of different orientations, (110) and
(100), and from different companies, \textit{MaTecK GmbH} and \textit{Metal
Crystals and Oxides Ltd.}, were studied. The \textit{MaTecK} crystals showed
lowest residual magnetic hysteresis and correspondingly clearer patterns of
the intermediate state. Figure \ref{xtalMH} shows magnetization loop measured
in a (100) - oriented Pb single crystal at $T=4.5$ K. The hysteresis vanishes
at $H\rightarrow0$ and minor hysteresis loops (shown by smaller open symbols
with the field sweep direction indicated by arrows) show no hysteresis. Larger
open symbols show result of a field-cooling experiment. They coincide with the
data obtained by sweeping magnetic field down. These observations implies zero
bulk pinning and we assert that the hysteresis comes from the difference in
topologies of the intermediate state upon flux entry and exit.

This assertion is directly confirmed by the magneto-optical images shown in
Fig. \ref{xtal}. Left column shows flux penetration, right column - flux exit.
There is obvious difference between the topologies of the flux patterns. The
tubes have sizes from $\mu m$ at low fields to sub-mm at higher fields. We
note that tubular structure is only observable in samples with no pinning
(pinning leads to dendrites \cite{prozorov2005}). Online vide-figures also
show robustness of the flux tubes upon penetration of a tilted field
\cite{video}. It is important to note that although laminar structure appears
on a hysteretic branch, it matches the field-cooled data indicating that this
irreversibility is not due to macroscopic flux gradient and cannot be
"quenched" by the annealing. We also note that tubular pattern appears on an
ascending branch that behaves as a textbook $M\left(  H\right)  $ loop,
providing additional evidence for its equilibrium state.%

\begin{figure}
[ptb]
\begin{center}
\includegraphics[
height=3.1756in,
width=3.8354in
]%
{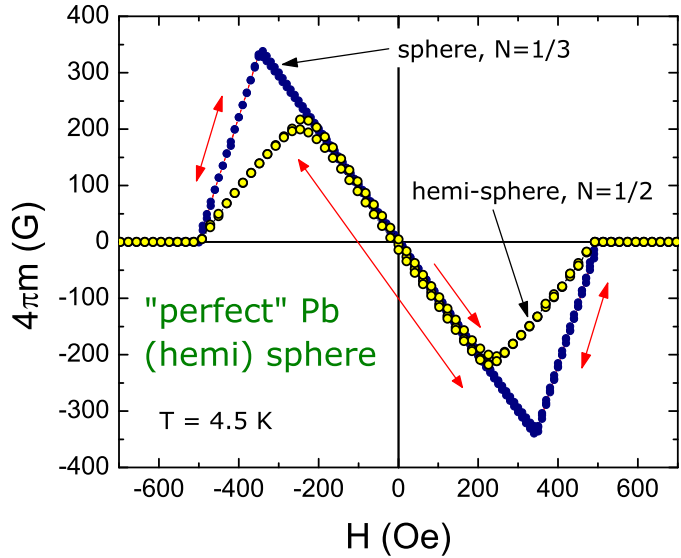}%
\caption{Magnetisation loop in a "perfect superconducting" sphere (solid
symbols) and a hemisphere (open symbols) at $T=4.5$ K.}%
\label{M(H)sphere}%
\end{center}
\end{figure}

\begin{figure}
[ptb]
\begin{center}
\includegraphics[
height=9.8299cm,
width=9.0457cm
]%
{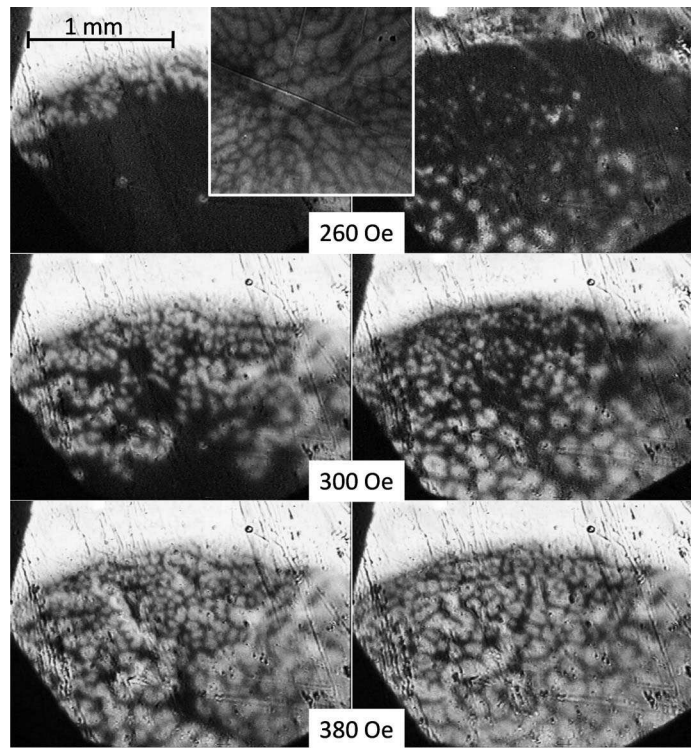}%
\caption{Flux penetration (left column) and exit (right column) in a Pb
hemisphere ($d=4$ mm) at $T=4$ K and indicated magnetic fields. Inset shows
tubular pattern in a different hemisphere sample.}%
\label{MOhemisphere}%
\end{center}
\end{figure}

In fact, observation of a closed topology on flux entry and open on flux exit
is quite typical for any sample with two flat surfaces perpendicular to the
applied field. The flux structure is governed by the geometric barrier
\cite{zeldov1994}. When a magnetic field is increased, Meissner currents flow
on both surfaces even in a thick disc \cite{prozorov2000}. Any closed-flux
object will be instantaneously driven to the sample center by the Lorentz
force. Therefore, it will appear as if the flux tubes pile up from the center
outwards, exactly as it is observed in Fig. \ref{xtal}. To eliminate the
geometric barrier the sample should be in form of an ellipsoid. In that case,
the Lorentz force is \textit{exactly} balanced by the condensation energy
force upon flux penetration. A comprehensive study of the shape effect on
magnetic hysteresis was reported in Ref. \cite{provost} where the authors
arrived to a clear conclusion that shape plays an important role and showed
that ellipsoidal sample has no hysteresis. Figure \ref{M(H)sphere} shows
experimentally such "perfect" $M\left(  H\right)  $ loops measured in a Pb
sphere (solid symbols) and a hemisphere (open symbols). The sphere was
produced by dropping molten lead (99.9999\% pure) in an inert atmosphere. The
hemisphere was cast into a copper mould in an inert atmosphere and
subsequently polished and annealed in vacuum at 250 $^{o}$C for 24 hours.
Still, the surface was not perfect and defects are seen in the imaging. Figure
\ref{MOhemisphere} shows the results for a hemisphere. (Inset shows another
hemisphere sample). The major result is evident - the geometric barrier is no
longer present (no dome-like formation of flux in the center) and flux tubes
appear \textit{both} ways - on flux entry and flux exit. Real - time video of
another hemisphere is available at \cite{video}.%

The video also shows flux tubes formation upon field cooling as well as
warming up from the pure Meissner state. Therefore, in a hemisphere flux tubes
appear from all four possible ways to reach a particular point inside the
intermediate state domain on an $H-T$ phase diagram.

\begin{figure}
[ptb]
\begin{center}
\includegraphics[
height=2.872in,
width=3.7014in
]%
{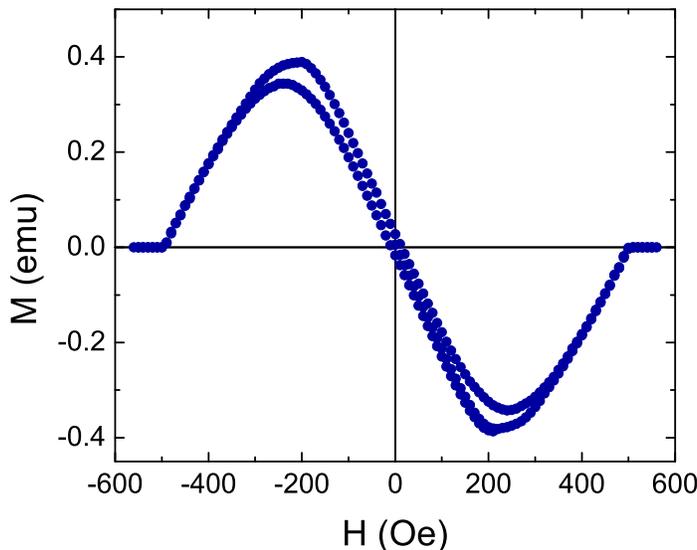}%
\caption{Magnetization loop in a cone at $T=4.5$ K.}%
\label{M(H)cone}%
\end{center}
\end{figure}

Another shape where geometric barrier is not present is a cone. The cones are
interesting, because one can go from an obtuse to an acute cone and quite
possibly the topology of the intermediate state will change. However, it is
very difficult to produce stress-free samples. We report data on the obtuse
cone (4 mm diameter, 1mm height) in Fig. \ref{M(H)cone}. The rounded shape of
the curve is apparently due to large local demagnetization at the corners.
Overall, the curve is quite reversible.%

\begin{figure}
[ptb]
\begin{center}
\includegraphics[
height=3.5578in,
width=3.5898in
]%
{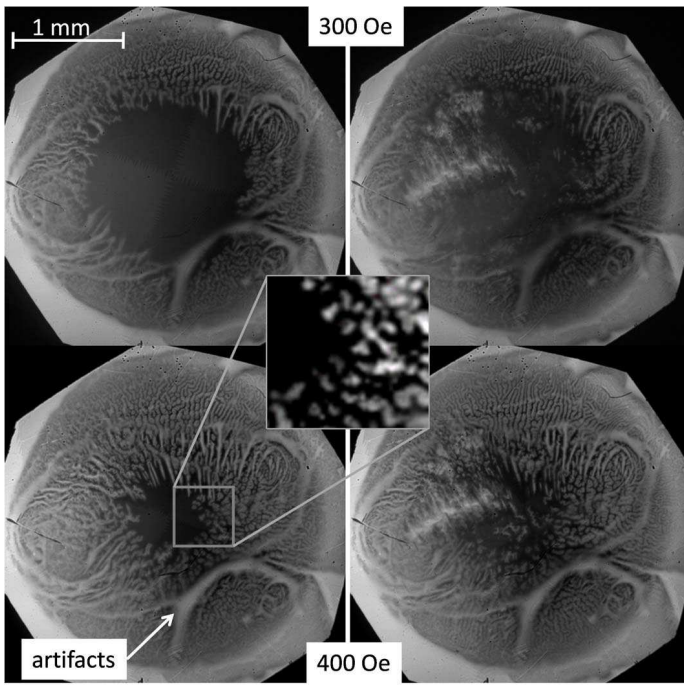}%
\caption{Flux penetration (left column) and exit (right column) into a
cone-shaped sample at $4$ K and at indicated fields. Inset shows a zoom of the
tubular pattern.}%
\label{cone}%
\end{center}
\end{figure}

Figure \ref{cone} shows magneto-optical imaging obtained in a conical sample.
Despite presence of some defects and artifacts (from grease and surface
imperfections), there is a clear absence of the geometric barrier and presence
of flux tubes.%

We conclude that tubular structure appears to be the equilibrium topology of
the intermediate state, because it is always observed on flux penetration in
samples of any shape, as well as on flux exit and upon warming and cooling in
constant field in samples without geometric barrier. Laminar structure on the
other hand is unstable in the presence of force - either Lorentz or
condensation energy. In flat samples, geometric barrier is present on
penetration, but not on exit. Combined with internal magnetic pressure for
flux to exit the latter preserves the laminar structure as metastable, but
flux-percolative state at high fields, which breaks into tubes at smaller
fields (as was foreseen by Landau \cite{landau1943}). In samples without
geometric barrier, there is always a condensation energy gradient that
destroys the laminar pattern. Our more recent experiments also show
transformation of the laminar pattern into tubular by applied external
current. Importantly, our observations emphasize that different topologies
result in actual macroscopic (topological) hysteresis in magnetization as
evident from Fig.\ref{xtalMH}.

I thank Jacob Hoberg and Baozhen Chen for help with the experiments.
Discussions with John Clem, Kotane Dam, Rudolf Huebener, Vladimir Kogan and
Roman Mints are greatly appreciated. Ames Laboratory is operated for the U.S.
Department of Energy by Iowa State University under Contract No.
W-7405-ENG-82. This work was supported by the NSF grant number DMR-05-53285,
the Alfred P. Sloan Foundation and by the Director for Energy Research, Office
of Basic Energy Sciences.

\end{document}